\documentclass[prl,twocolumn,showpacs,preprintnumbers,amsmath,amssymb]{revtex4}

\usepackage{graphicx}
\usepackage{dcolumn}
\usepackage{bm}

\begin{document}

\preprint{APS/123-QED}

\title{Universal Scaling Behavior of Anomalous Hall Effect and Anomalous 
Nernst Effect in Itinerant Ferromagnets}

\author{T. Miyasato$^1$, N. Abe$^1$, T. Fujii$^1$, A. Asamitsu$^{1,2}$, 
S.Onoda$^2$, Y. Onose$^3$, N. Nagaosa$^{2,3,4}$ and Y. Tokura$^{2,3,4}$
 }

\affiliation{%
$^1$Cryogenic Research Center, University of Tokyo, Tokyo 113-0032, Japan \\
$^2$Spin Superstructure Project, ERATO, JST, AIST Central 4, Tsukuba 305-8562, Japan\\
$^3$Department of Applied Physics, University of Tokyo, Tokyo 113-8656, Japan \\
$^4$Correlated Electron Research Center(CERC), AIST Central 4, Tsukuba 305-8562, Japan
}%
\date{\today}

\begin{abstract}
 Anomalous Hall effect (AHE) and anomalous Nernst effect (ANE) in 
a variety of ferromagnetic metals including pure metals, oxides, and chalcogenides, are studied to obtain unified understandings of their origins. We show a universal scaling behavior of anomalous Hall conductivity 
$\sigma_{xy}$ as a function of longitudinal conductivity $\sigma_{xx}$ 
over five orders of magnitude, which is well explained by a recent theory 
of the AHE taking into account both the intrinsic and extrinsic contributions. 
ANE is closely related with AHE and provides us with further information about 
the low-temperature electronic state of itinerant ferromagnets. 
Temperature dependence of transverse Peltier coefficient $\alpha_{xy}$ 
shows an almost similar behavior among various ferromagnets, and 
this behavior is in good agreement quantitatively with that expected from the 
Mott rule.
\end{abstract}

\maketitle

It has been known that Hall resistivity $\rho_{yx}$ in an itinerant ferromagnet 
has an extra contribution from spontaneous magnetization $M$, which 
is often expressed empirically by the formula $\rho_{yx}=R_0 H + 4\pi R_S M$, 
where $R_0$ and $R_S$ denote ordinary and anomalous Hall coefficient, respectively, and $R_S$ 
is usually a function of the resistivity of materials\cite{label1}. The origin
 of anomalous Hall effect (AHE), however, has long been an 
intriguing but controversial issue since 1950s. Some of the theories explain 
 AHE  from extrinsic origins such as skew scattering 
($\rho_{yx}\propto \rho_{xx}$)\cite{label2} or side-jump 
($\rho_{yx} \propto \rho_{xx}^2$)\cite{label3} mechanisms 
due to the spin-orbit interaction. \\
\indent
In contrast to these extrinsic mechanisms, several works point out the
 intrinsic origin of the AHE, which is closely related to the quantal Berry
 phase on Bloch electrons in solids\cite{label4,label5,label6,label7,label8}:
 the intrinsic part of the anomalous Hall conductivity is given by the sum of
 the Berry-phase curvature of the Bloch wavefunction over the occupied states,
 in an analogy to the quantum Hall effect. This Berry-phase scenario of the AHE
 has recently attracted much interest for its dissipationless and topological
 nature. Using first-principle band calculations, the intrinsic anomalous Hall
 conductivity has been calculated for ferromagnetic semiconductors\cite{label8,
 label9}, transition metals\cite{label10, label11, label12}, and 
oxides\cite{label13,label14}, in quantitative agreement with experimental
 results. For example, the AHE in ruthenates (SrRuO$_3$) was found to be very sensitive to
 details of the electronic band structure such as the location of (nearly)
 crossing points of band dispersions. Such a momentum point acts 
as a ``magnetic monopole'' yielding a large Berry-phase curvature and 
resulting in a resonant enhancement of the anomalous Hall 
conductivity\cite{label13}.\\
\indent
Recently, a theory of AHE has been developed taking into account this 
resonant contribution from the band crossing, where both the 
topological dissipationless current and dissipative transport current 
are treated in the presence of the impurity scattering in a unified 
way\cite{label15}. It proposes three scaling regimes for the AHE as a 
function of the electron lifetime or the resistivity. \\
\indent
In the present paper, we report the anomalous Hall effect in a variety 
of itinerant ferromagnets at low temperatures, where  
the magnetization of materials is almost saturated. We have performed 
AHE measurement on pure metals (Fe, Co, Ni, and Gd films), 
oxides(SrRuO$_3$ crystal(SRO)\cite{label13}, 
La$_{1-x}$Sr$_x$CoO$_3$ crystals(LSCoO)\cite{label16}), and 
chalcogenide-spinel crystals(Cu$_{1-x}$Zn$_x$Cr$_2$Se$_4$)\cite{Lee_science04}. 
We have found a universal scaling behavior of the transverse conductivity 
$\sigma_{xy}$ as 
a function of longitudinal conductivity $\sigma_{xx}$, which is in 
good quantitative agreement with a  unified theory of the AHE taking 
into account  both intrinsic and extrinsic origins\cite{label15}. 
In addition to AHE, we have also examined anomalous Nernst effect 
(ANE, transverse thermoelectric effect in the presence of spontaneous
 magnetization), which will provide us with another useful information on 
the electronic ground state and its relation to the Berry-phase scenario on 
the AHE.\\
\begin{figure}
\includegraphics[width=82mm,height=160mm,keepaspectratio, clip]{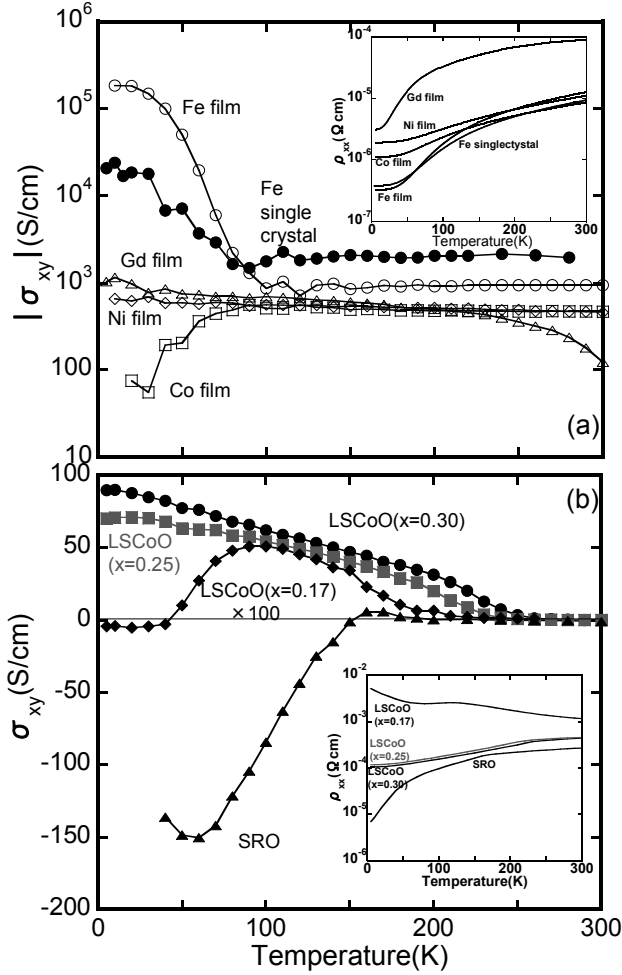}
\caption{\label{figure1} Temperature dependence of anomalous Hall conductivity  $\sigma_{xy}$ for (a) pure metals  
in logarithmic scale and (b) oxides  in linear scale. In top panel, 
$\sigma_{xy}$ is negative quantity for Ni and Gd. Temperature dependence of 
longitudinal resistivity $\rho_{xx}$ is also shown in the inset.}

\end{figure}
\indent
We used thin films of Fe, Co, Ni and Gd with the thickness of $1\mu $m and the purity of 99.85$\%$, 99.9$\%$, 99+$\%$, 99.9$\%$, respectively. Single crystals of  La$_{1-x}$Sr$_x$CoO$_3$ (x=0.17, 0.25 and 0.30) and SrRuO$_3$ were grown 
by a floating-zone method and a flux method, and whose Curie temperatures are 120K, 225K, 235K, and 160K, respectively. 
The Hall resistivity $\rho_{yx}$ was measured using a Physical Properties Measurement System (Quantum Design Co., Ltd.)  together with the longitudinal resistivity $\rho_{xx}$ as a function of magnetic field ($H$) and 
temperature ($T$). 
The transverse thermopower Q$_{yx}=E_y /\partial_x T$ was measured using the same platform by introducing necessary wirings. We apply a temperature gradient $\partial_x T$ to an electrically isolated sample in a magnetic field and measure the transverse voltage appeared. 
According to the linear transport theory, we have 
$\vec{j} = \tilde{\sigma} \vec{E} + \tilde{\alpha} (-\nabla T)$ and 
$\vec{E} = \tilde{\rho} \vec{j} + \tilde{Q} \nabla T$, 
where $\vec{j}$ stands for the electric current and $\vec{E}$ for the electric 
field, and $\tilde{\sigma}$, $\tilde{\rho}$, $\tilde{\alpha}$, and 
$\tilde{Q}$ denote conductivity-, resistivity-, Peltier-, and thermoelectric 
tensors, respectively. 
Then we obtain $Q_{xy}=-E_y/\partial_x T+Q_{xx}(\partial_y T/\partial_x T)$ as 
$\vec{j}=0$. 
Because we confirmed that the second term is negligibly small compared to the first term, we defined $Q_{xy}=-E_y/\partial_x T$ in the following.  
The anomalous contribution in $\rho_{yx}$ and $Q_{xy}$ was determined by 
extrapolating $\rho_{yx}$ and Q$_{xy}$ vs. $H$ curves to $H=0$. The transverse conductivity $\sigma_{xy}$ was estimated as $-\rho_{xy}/\rho_{xx}^2$ and transverse Peltier coefficient $\alpha_{xy}$ as 
$(Q_{xy}-Q_{xx}\tan\theta _{xy})/\rho_{xx}$, where 
$\theta_{xy}$ being the Hall angle. 
The contribution from magnetoresistance or magnetothermopower was carefully 
 removed by subtracting $\rho_{yx}(-H)$ from $\rho_{yx}(H)$ or 
$Q_{xy}(-H)$ from $Q_{xy}(H)$. 

Figure 1 shows the temperature dependence of anomalous Hall conductivity 
$\sigma_{xy}$ in pure metals(upper panel, (a)) and oxides(bottom panel, (b)).
 Note that the scale of vertical axis in Fig.1 (a) is 
logarithmic of $|\sigma_{xy}|$  while it is linear in Fig.1 (b). 
All the ferromagnets studied are metallic except LSCoO ($x=0.17$) as 
seen in the inset of Fig.1. In pure metals, the value of $|\sigma_{xy}|$ is 
almost constant with $100-1000$ S/cm below room temperature down to 100K, 
and then varies significantly for Fe and Co down to absolute zero. The 
magnetization $M$ in pure metals is constant below room temperature 
(not shown). In oxides, the change of $\sigma_{xy}$ is very complicated due to 
the change of $M$ and even shows the sign change in SRO and LSCoO. 

A striking relation among various $\sigma_{xy}$ values becomes apparent
if we focus on $\sigma_{xy}$ at a lowest temperature. 
Figure 2 shows 
the variation of the absolute value of anomalous Hall conductivity 
$|\sigma_{xy}|$ against the longitudinal conductivity $\sigma_{xx}$ over 
five orders of magnitude in the ground state of itinerant ferromagnets. 
The data for Cu$_{1-x}$Zn$_x$Cr$_2$Se$_4$ is included in the figure. 
For pure metals, all data of $|\sigma_{xy}|$ below room temperature are 
also plotted.
The variation of $|\sigma_{xy}|$ can be categorized into three regions: 
In the intermediate region with $\sigma_{xx}=10^4 - 10^6$ S/cm, 
such as in pure metals and SrRuO$_3$, one can see that $|\sigma_{xy}|$ is nearly constant 
($\simeq 1000$ S/cm), which means that $\rho_{yx} \propto \rho_{xx}^2$. Furthermore, this constant value of $|\sigma_{xy}|$ is consistent with the ``resonant'' AHE which gives the intrinsic contribution of the order of $e^2/ha \sim 10^3$ S/cm, with $a$ being a lattice constant\cite{label15}. The contributions from the extrinsic mechanisms, i.e., skew-scattering and side-jump, are found to be much smaller than $e^2/ha$ in this region. Therefore, we can regard the $\sigma_{xy}$ in the plateau region as the dominantly 
intrinsic contribution.

In the extremely clean case with $\sigma_{xx} \simeq 10^6$ S/cm, such as in  Fe and Co at low temperatures, the behavior of  $|\sigma_{xy}|$ seems to depend on materials. According to the classical Boltzmann transport theory, impurity scattering gives rise to anomalous Hall conductivity through the skewness or the side jump, and the skew-scattering contribution to AHE diverges in the clean limit as $\sigma_{xy}\propto \sigma_{xx}$. Although the experimental results show a slight deviation from the theoretical 
prediction, the qualitative change in $\sigma_{xy}$ from the intrinsic region is obvious. 

Finally in the dirty limit with $\sigma_{xx}<10^4$ S/cm, such as in 
Cu$_{1-x}$Zn$_x$Cr$_2$Se$_4$ and La$_{1-x}$Sr$_x$CoO$_3$, the intrinsic contribution to AHE is suppressed by the damping effect due to impurities, and the change in anomalous Hall conductivity is well described by $\sigma_{xy}\propto \sigma_{xx}^{1.6}$ experimentally. This exponent is also expected in the ``insulator'' regime of quantum-Hall systems\cite{Pryadko}.

\begin{figure}
\includegraphics[width=88mm,height=160mm,keepaspectratio, clip]{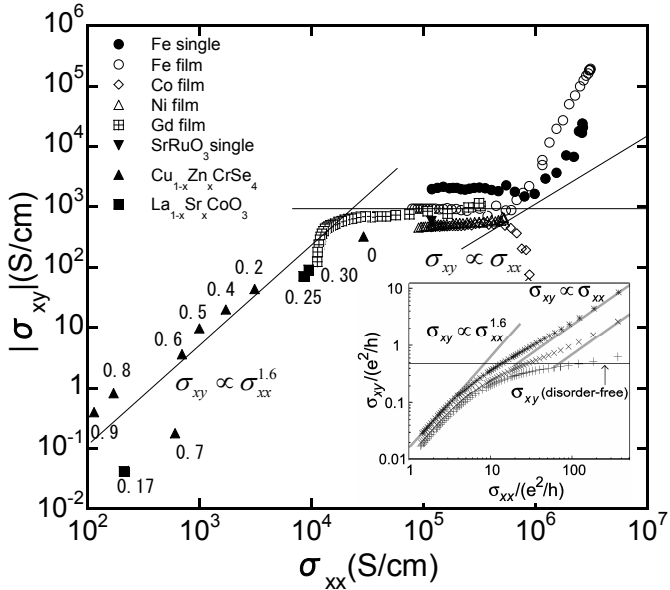}
\caption{\label{figure2} Absolute value of anomalous Hall  conductivity 
$|\sigma_{xy}|$ as a function of longitudinal conductivity $\sigma_{xx}$ in pure metals(Fe, Ni, Co, and Gd), oxides(SrRuO$_3$ and La$_{1-x}$Sr$_x$CoO$_3$), and chalcogenide spinels (Cu$_{1-x}$Zn$_x$Cr$_2$Se$_4$) at low temperatures. Three lines are $\sigma_{xy}\propto \sigma_{xx}^{1.6}$, 
$\sigma_{xy}=$const., and $\sigma_{xy}\propto \sigma_{xx}$ for dirty, intermediate, and clean regimes, respectively. The inset shows theoretical results obtained from the same analysis as in Fig. 4 of Ref.[15] but for 
$E_F=0.9$; $2mv=0.02$ ($+$), 0.2 ($\times$), and 0.6 ($*$). Here, $m$ is the effective mass and $v$ is the strength of the $\delta$-functional impurity potential. 
}
\end{figure}

The universal scaling behavior above is well explained 
by a unified theory of the AHE taking into account 
both intrinsic and extrinsic origins\cite{label15}:
Three scaling regimes have been found for a generic two-dimensional model containing the resonant enhancement of $\sigma_{xy}$ due to an anti-crossing of band dispersions and the impurity scattering.
In the extremely clean case where the relaxation rate $\tau^{-1}$ is smaller than the band energy splitting given by the spin-orbit interaction 
energy $\varepsilon_{so}$, the extrinsic skew-scattering contribution gives the leading contribution, yielding the scaling $\sigma_{xy}\propto\sigma_{xx}$. 
If the Fermi level is located around the anti-crossing of band dispersions, a crossover to the intrinsic regime occurs around $\tau^{-1}\sim\varepsilon_{so}$, with the resonant enhancement $\sigma_{xy}\sim e^2/h$ and the scaling $\sigma_{xy}=\text{constant}$. For the hopping-conduction regime with $\tau^{-1}>E_F$ with the Fermi Energy $E_F$, there occurs another scaling $\sigma_{xy}\propto\sigma_{xx}^{1.6}$.
The present experimental results on the crossover in $\sigma_{xy}$ among clean, intermediate, and dirty cases is thus well reproduced by this theory (see 
the inset of Fig.2). 

\begin{figure}
\includegraphics[width=85mm,height=160mm,keepaspectratio, clip]{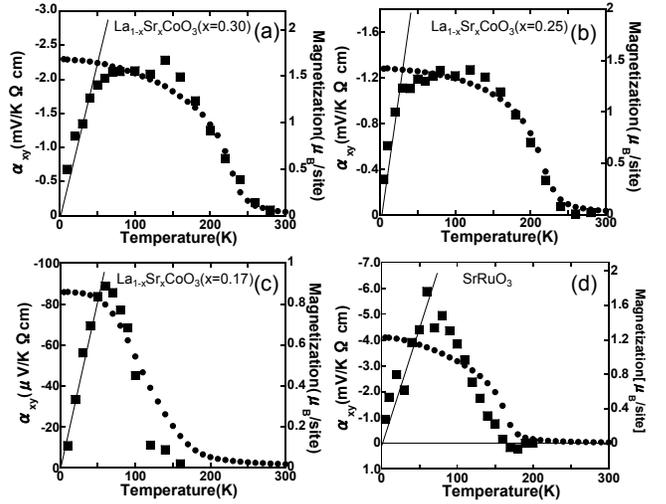}
\caption{\label{figure3}Temperature dependence of anomalous transverse Peltier coefficient $\alpha_{xy}$ (squares) and magnetization $M$ (circles) in 
(a) La$_{1-x}$Sr$_{x}$CoO$_3$ ($x=0.3$), (b)La$_{1-x}$Sr$_{x}$CoO$_3$ ($x=0.25$), (c) La$_{1-x}$Sr$_{x}$CoO$_3$ ($x=0.17$), and (d) SrRuO$_3$. 
The straight line at low temperatures represents $T$-linear variation of 
$\alpha_{xy}$.}
\end{figure}

\begin{table*}
\caption{\label{table1} Experimental data of $\alpha_{xy}/T$, 
$\sigma_{xy}$ at a lowest temperature, carrier density $n$ derived from 
the ordinary 
Hall effect at room temperature, and electronic heat-capacity coefficient 
$\gamma$ in La$_{1-x}$Sr$_x$CoO$_3$.
The left-hand side (LHS) in Eq.(2) is the average of 
two successive $\alpha_{xy}/T$ 
and the right-hand side (RHS) is estimated using the average of $\gamma$ 
and the difference $\Delta \sigma_{xy}/\Delta n$. 
}
\begin{ruledtabular}
\begin{tabular}{cc|c||c|ccc}
 $x$ & $\alpha_{xy}/T$($\mu$V/K$^2$$\Omega$cm) 
& LHS & RHS & $\sigma_{xy}$(S/cm) & 
$n$(10$^{23}$/mol) & $\gamma$(mJ/mol K$^2$)  \\ 
\hline
 0.30&-437&&&90.3 &1.43 &49.1  \\ 
     &&-285&-339&\\
 0.26&-132&&&34.2 &1.89 &39.5  \\
     &&-142&-170&\\
 0.22&-153&&&23.8 &2.04 &41.1  \\
     &&-85.9&-103&\\
 0.18-2&-19.2&&&1.10 &2.86 &19.0  \\
     &&-27.8&-8.77&\\
 0.18-1&-8.62&&&-0.002 &3.27 &32.4  \\
\end{tabular}
\end{ruledtabular}
\end{table*}

 Now we move on to the anomalous Nernst effect.  The transverse Peltier coefficient $\alpha_{xy}$ is given by the Mott rule,
\begin{eqnarray}
\alpha_{xy}= \left(\frac{\pi ^2k_B^2}{3e} \right)T\frac{d}{d\epsilon }\left[\sigma _{xy} \left(\epsilon \right) \right]_{\mu}\label{4}, 
\end{eqnarray}
where $k_B$ is the Boltzmann constant, $e$ the elementary charge, and 
$\mu$ the chemical potential\cite{label17}.  In Fig.3, we show $\alpha_{xy}$ and $M$ simultaneously in La$_{1-x}$Sr$_{x}$CoO$_3$(x=0.3, 0.25, 0.17) and
SrRuO$_3$. All the materials (not shown for pure metals) seem to show qualitatively very similar temperature dependence  that  $\alpha_{xy}$ starts to increase just below $T_C$, being almost proportional to $M$, then decreases at low temperatures linearly with $T$, and finally vanishes toward absolute zero. 
These behaviors are well understood using the above formula: 
$\alpha_{xy}$ just below $T_C$ is subject to the factor 
($d \sigma _{xy}/d\epsilon)_{\mu}$, 
 where the modification of the band structures at the Fermi level 
takes place due to the ferromagnetic transition. After the saturation of $M$, 
$T$-linear term becomes dominant in the change of $\alpha_{xy}$. 

In order to confirm the validity of Eq.(1) further, we performed a quantitative analysis on $\alpha_{xy}$ in 
LSCoO(x=0.3$-$0.18). The equation is rewritten to 
\begin{eqnarray}
\frac{\alpha_{xy}}{T}= \frac{\gamma}{e}\frac{d}{dn}\left[\sigma _{xy} \left(\epsilon \right) \right]_{\mu}\label{5}, 
\end{eqnarray}
with the electronic heat-capacity coefficient $\gamma$  by using 
the transformation 
$\frac{d}{d\epsilon }\left[\sigma _{xy} \left(\epsilon \right) \right]_{\mu}
=\frac{dn}{d\epsilon }\frac{d}{dn}\left[\sigma _{xy} \left(\epsilon \right) \right]_{\mu}$. 
We obtained $\gamma$ from heat-capacity measurement and the 
carrier density $n$ from the ordinary 
Hall effect at 300K which is far above $T_{C}$. 
These values are shown in Table \ref{table1}. 
Because the composition $x$ is nominal, we employed two samples with 
$x=0.18$ showing different values of physical quantities\cite{note1}.  
The left-hand and right-hand sides (LHS and RHS) of Eq.(\ref{5}) were estimated 
independently using these values in the center of Table \ref{table1}, 
replacing the differential $d\sigma_{xy}/dn$ 
with the difference $\Delta \sigma_{xy}/\Delta n$, and the relation of both 
sides was summarized in Fig.4. 
The proportionality between these two quantities is obvious, and the slope is about 0.85 and slightly different from 1. Although we do not understand 
the reason for this slight discrepancy yet, $\alpha_{xy}$ obeys  Eq.(2) 
 quantitatively, and hence the thermoelectric Hall transport property is 
understandable, at least in the dirty case, in terms of the Mott rule.
This may suggest that 
the Berry-phase contribution to the thermoelectric Hall 
transport phenomena is dominant mainly through $\sigma_{xy}$.

\begin{figure}
\includegraphics[width=77mm,height=160mm,keepaspectratio, clip]{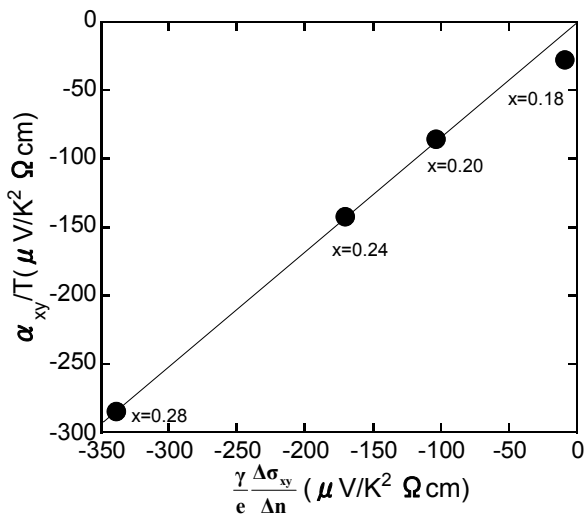}
\caption{\label{figure4}
The relation between the left-hand and 
right-hand sides of Eq.(2) derived independently using the 
experimental data in Table \ref{table1} for La$_{1-x}$Sr$_x$CoO$_3$. 
The line has a gradient of about 0.85. }
\end{figure}

In summary, we have investigated the anomalous Hall effect and anomalous 
Nernst effect in various ferromagnetic metals, such as Fe , Co, Ni, Gd, La$_{1-x}$Sr$_x$CoO$_3$, SrRuO$_3$, and Cu$_{1-x}$Zn$_x$Cr$_2$Se$_4$. The anomalous Hall conductivity 
$\sigma_{xy}$ in the ground state  shows a universal scaling behavior 
against the longitudinal conductivity 
$\sigma_{xx}$, being independent of 
materials. This scaling relation  can be well understood by a recent  
theory taking into account both intrinsic and extrinsic origin of the AHE. 
We have also shown that the relation between the anomalous Nernst effect 
and the anomalous Hall effect can be explained  quantitatively by the Mott 
rule. 

This work was partly supported by the Grant-in-Aid for 
Scientific Research (Nos.15104006, 16076205, 17105002, and 17038007) from 
the Ministry of Education, Culture, Sports, Science and Technology, Japan.


\begin{thebibliography}{00}
\bibitem{note1}
  This doping level is close to the metal-insulator boundary in 
La$_{1-x}$Sr$_x$CoO$_3$\cite{label16}, and hence the propeties are quite 
sensitive to slight off-stoichiometry.
\end{thebibliography}
\end{document}